\documentstyle[aps,12pt]{revtex}

\begin{document}
\draft
\title{Comment on ''Understanding the Area Proposal for Extremal Black Hole
Entropy''}
\author{O. B. Zaslavskii}
\address{Department of Physics, Kharkov State University, Svobody Sq.4, Kharkov\\
310077, Ukraine\\
E-mail: olegzasl@apt.kharkov.ua}
\maketitle

\begin{abstract}
A. Ghosh and P. Mitra made the proposal how to explain the area law for the
entropy of extreme black holes in some model calculations. I argue that
their approach implicitly operates with strongly singular geometries and
says nothing about the contribution of regular metrics of extreme black
holes into the partition function.
\end{abstract}





In the recent paper \cite{Und} the attempt has been made to explain the area
law for extreme black holes. The authors write down the partition function
with account for topologies of either extreme or nonextreme black holes: $%
Z=\Sigma _{\text{topologies}}\int d\mu (m)\int d\mu (q)e^{-I(m,q)}$ where $m$
is a mass, $q$ is an electric charge. They point out that the action $I$ has
different form for nonextreme ($I_{n}$) and extreme ($I_{e}$) cases: 
\begin{equation}
I_{n}=\beta (m-q\Phi )-\pi (m+\sqrt{m^{2}-q^{2}})^{2},I_{e}=\beta (m-q\Phi )
\label{2}
\end{equation}
where $\beta $ is the inverse temperature on a boundary (which is supposed
in \cite{Und} to be situated at infinity), $\Phi $ is the difference of
potentials between a boundary and horizon. The next step in derivation the
area law in \cite{Und} is crucial. It consists in the statement that ''the
nonextremal action is lower than the extremal one {\it for each set of
values of }$q,m$'' (italic is mine - O.Z.) In other words, for fixed $\beta $
and $\Phi $ 
\begin{equation}
I_{n}(m,q)<I_{e}(m,q)  \label{3}
\end{equation}
From eq.(\ref{3}) authors conclude that it is the nonextreme topology which
mainly contributes to $Z$ whence they ''explain'' the area law for the
entropy.

However, eq.(\ref{3}) implies that $m$ and $q$ in both sides of it are the
same. As for nonextreme black holes $m\neq q$ it must hold also for
solutions called in \cite{Und} ''extreme'' and whose action is calculated
above with the zero entropy term. As I show below this leads to singular
behavior of the metric. The metric of a spherically-symmetrical charged
black hole can be written as 
\begin{equation}
ds^{2}=b^{2}d\tau ^{2}+\alpha ^{2}dr^{2}+r^{2}d\Omega ^{2}  \label{5}
\end{equation}
It follows from the Gauss law and Hamiltonian constraint \cite{RN} that 
\begin{equation}
\alpha ^{2}=(1-\frac{2m}{r}+\frac{q^{2}}{r^{2}})^{-1}  \label{6}
\end{equation}
If $m\neq q$ this entails the asymptotics $\alpha ^{2}\propto (r-r_{+})^{-1}$
near an event horizon at $r=r_{+}.$ On the other hand, to distinguish
extreme and nonextreme topologies it was used in \cite{Und} that $\xi \equiv
\lim_{r\rightarrow r_{+}}b^{\prime }/\alpha $ $=0$ in the first case in
contrast to $\xi =1$ in the second one. It means that instead of $%
b^{2}\propto (r-r_{+})$ typical of nonextreme holes this coefficient behaves
in the extreme case as $b^{2}\propto (r-r_{+})^{1+n}$ with $n>0.$

However, both asymptotics $\alpha ^{2}\propto (r-r_{+})^{-1}$ and $%
b^{2}\propto (r-r_{+})^{1+n}$ are inconsistent with each other in that the
Riemann curvature diverges strongly at the horizon either in Euclidean or
Lorentzian version:\thinspace $R\propto l^{-2}$ where $l$ is a proper
distance from the horizon. Although integrals from $R$ entering into the
action converge, such behavior of geometry looks unphysical. Moreover, all
other curvature invariant diverge as well. For example, $R_{\mu \nu \sigma
\rho }R^{\mu \nu \sigma \rho }$ $\propto l^{-4}.$ In other words, the
''extreme'' solutions discussed in \cite{Und} correspond to spacetimes for
which the whole surface $r=r_{+}$ consists of singular points. It follows
from the above discussion that solutions at hand do not compose some
particular, very special class of configurations which may or may not be
used in $Z$ at one's will. On the contrary, the properties described above
hold true for {\it any }solution with $m\neq q$ and $\xi =0.$ One may ask,
whether all that means that the result of \cite{Und} consists in the proof
that the true physical equilibrium cannot be achieved at singular geometries
discussed above. This would restrict the significance of the result since it
gives no information about the relative contribution of regular extreme
topologies but, at least, it would mean that it can be given reasonable
interpretation. Unfortunately, even in this restricted sense the ''proof''
presented in \cite{Und} is incorrect as is argued below.

Let me recall the main points which prescription for the entropy $S=0$ and
temperature ($\beta $ is arbitrary$)$ of extreme black holes are based on 
\cite{Extr}: (i) the condition of the regularity of Euclidean geometry which
is satisfied for any $\beta $ (for nonextreme holes (i) is the same but is
satisfied by the only choice of $\beta $); (ii) topological meanings, in
which the crucial role is played by the property $l=\infty $ between a
horizon and any other point outside it. {\it Both (i) and (ii) are not
satisfied by ''extreme'' black holes with }$m\neq q.${\it \ }As far as (i)
is concerned, the situation is completely reverse to that in \cite{Extr}:
whereas the Euclidean geometry in \cite{Extr} is {\it regular for any }$%
\beta $, the Euclidean geometry (as well Lorentzian one) in \cite{Und} is%
{\it \ singular for any }$\beta .$ As regards (ii), $l$ is finite for any $%
m\neq q$. In view of motivation for $S=0$ fails for solutions under
discussion, the eq.(\ref{3}) loses its meaning at all (even if singular
solutions are admitted) that destroys the ''proof'' completely. I stress
that this takes place irrespectively of how small the difference $m-q$ is.

{\it The crucial point in which the analysis of \cite{Und} flaws consists in
dealing with metrics for which simultaneously }$\xi =0${\it \ and }$m\neq q.$
As a result, ''explanation'' of the area proposal for extreme black holes in 
\cite{Und} is unsound.

The precedent text is published as Comment \cite{zasl98}. Below I give a
brief response to Ghosh and Mitra's Reply \cite{reply}.

1) Ghosh and Mitra state that Eq. (1) ''was neither written nor used in \cite
{Und}''. This inequality was not, indeed, written explicitly in \cite{Und}
as a separate formula (probably, because of its simplicitly). However, it
was not only used by Ghosh and Mitra - it is the key point in their
''proof'': ''It is clear from (7) that the nonextremal action is lower than
the extremal one for each set of values of $q,m$. Consequently, the
partition function is to be approximated by $e^{-I_{q,m}}$(...)'' (The
paragraph after Eq. (8) in \cite{Und}). Here $I_{q,m}=I_{n}(m,q)$ in my
notations, Eq. (7) of \cite{Und} corresponds to \ref{2} above. As a matter
of fact, Ghosh and Mitra repeat (\ref{3}) in Reply in a slightly different
form: 
\begin{equation}
I_{n}(m,q-\varepsilon )<I_{e}(m,q)  \label{7}
\end{equation}
with small but nonzero $\varepsilon .$ In fact, however, the quantity $I_{n}$
is the continuous function of its arguments, so one may take the limit $%
\varepsilon \rightarrow 0$ safely and there is no essential difference
between (\ref{7}) and (\ref{2}). Therefore, all arguments pushed forward as
regards \ref{3} retain their validity with respect to \ref{7}. That charges
in both sides of (\ref{7}) may slightly different from each other, is not
crucial. Much more important is that in the r.h.s. of (\ref{7}) $m$ and $q$
may be different and this leads to difficulties indicated in Comment.

2) It is demonstrated explicitly in \cite{zasl98} that {\it every}
configuration among those discussed in \cite{Und} leads to a singular
geometry. Authors of \cite{reply} keep silence with respect to the
corresponding arguments, thus continuing to state that my objection concerns
only some particular classes of metrics (calling the crucial point ''the
technical observation'') without any refutation of the claim made in \cite
{zasl98} about the generality of the property under discussion. Instead of
it, Ghosh and Mitra emphasize: ''these configurations were not explicitly
used by us, nor do they need to be used implicitly'' as if one might discard
at any moment some metrics (in fact, all of them with $m\neq q$ in our case)
from the path integral at his own will.

3) Inasmuch as the action integral converges, the singular behavior of the
metrics under discussion does not bother authors of \cite{Und}, \cite{reply}
Such a liberal attitude to singularities conflicts with what is implied to
be done with them. For instance, in the conical singularity approach for
nonextreme black holes the action integral certainly converges since the
singularity even is weaker than that in \cite{Und} (delta-like instead of
power-like). In spite of it, this singularity is not accepted physically.
For higher-curvature Lagrangians the situation becomes worse since in that
case even the action integral would diverge under conditions considered in 
\cite{Und}. It is worth noting that the demand of regularity near the
horizon imposes general restrictions on a black hole metric \cite{Constr}.
Metrics considered in \cite{Und} do not obey the corresponding conditions.

The last claim of \cite{reply} that the action integral is calculated
directly without any additional assumptions seems to be correct. However,
again, such a strange object as an ''extreme'' black hole with a finite
proper distance between any point and the ''horizon'' and with the entropy $%
S=0$ is gained due to dropping such a fundamental condition as the
regularity of a manifold and, therefore, hardly has any physical sense.

In general, three different topological sectors are involved into
competition for given boundary data $(\beta ,\Phi )$: a flat geometry,
regular nonextreme metrics with $m\neq q$ and regular extreme ones with $m=q$%
. The paper \cite{Und} says nothing, however, about their relative
contribution and deals with singular ''extreme'' configurations for which $%
m\neq q$.

Let me conclude with some general remarks beyond the issues discussed in 
\cite{zasl98} and \cite{reply}. It is quite possible that any estimate of
different fractional contributions to the partition function made in the
zero- loop approximation (even for regular geometries) will be unphysical.
The point is that the temperature of extreme black holes, according to \cite
{Extr}, is a finite nonzero quantity, i.e. it differs from the Hawking one.
This assumption leads to the divergencies of the stress-energy tensor on an
event horizon of a generic extreme black hole \cite{And}. Although this
result is obtained in the semiclassical approximation whereas the geometry
in the situation considered in \cite{Und} fluctuates itself, it is very
likely that path integral over geometries each of which is singular itself
will diverge. In other words, quantum effects can change the semiclassical
thermodynamics of extreme black holes drastically.

\end{document}